\documentclass[twocolumn,preprintnumbers,amssymb,amsmath,superscriptaddress,nofootinbib]{revtex4}
\usepackage{graphicx}
\usepackage{dcolumn}
\usepackage{bm}
\usepackage{comment}
\begin{document} 
\input epsf.tex
\newcommand{\beq}{\begin{eqnarray}}
\newcommand{\eeq}{\end{eqnarray}}
\newcommand{\nn}{\nonumber}
\def\ltap{\ \raise.3ex\hbox{$<$\kern-.75em\lower1ex\hbox{$\sim$}}\ }
\def\gtap{\ \raise.3ex\hbox{$>$\kern-.75em\lower1ex\hbox{$\sim$}}\ }
\def\CO{{\cal O}}
\def\CL{{\cal L}}
\def\CM{{\cal M}}
\def\tr{{\rm\ Tr}}
\def\CO{{\cal O}}
\def\CL{{\cal L}}
\def\CM{{\cal M}}
\def\mpl{M_{\rm Pl}}
\newcommand{\bel}[1]{\be\label{#1}}
\def\al{\alpha}
\def\bt{\beta}
\def\eps{\epsilon}
\def\eg{{\it e.g.}}
\def\ie{{\it i.e.}}
\def\mn{{\mu\nu}}
\newcommand{\rep}[1]{{\bf #1}}
\def\be{\begin{equation}}
\def\ee{\end{equation}}
\def\bea{\begin{eqnarray}}
\def\eea{\end{eqnarray}}
\newcommand{\eref}[1]{(\ref{#1})}
\newcommand{\Eref}[1]{Eq.~(\ref{#1})}
\newcommand{\gsim}{ \mathop{}_{\textstyle \sim}^{\textstyle >} }
\newcommand{\lsim}{ \mathop{}_{\textstyle \sim}^{\textstyle <} }
\newcommand{\vev}[1]{ \left\langle {#1} \right\rangle }
\newcommand{\bra}[1]{ \langle {#1} | }
\newcommand{\ket}[1]{ | {#1} \rangle }
\newcommand{\ev}{{\rm eV}}
\newcommand{\kev}{{\rm keV}}
\newcommand{\Mev}{{\rm MeV}}
\newcommand{\gev}{{\rm GeV}}
\newcommand{\tev}{{\rm TeV}}
\newcommand{\mev}{{\rm MeV}}
\newcommand{\meV}{{\rm meV}}
\newcommand{\mnu}{\ensuremath{m_\nu}}
\newcommand{\nnu}{\ensuremath{n_\nu}}
\newcommand{\mlr}{\ensuremath{m_{lr}}}
\newcommand{\acc}{\ensuremath{{\cal A}}}
\newcommand{\mav}{MaVaNs}
\newcommand{\disc}[1]{{\bf #1}} 
\newcommand{\mh}{{m_h}}
\newcommand{\hb}{{\cal \bar H}}
\newcommand{\me}{\mbox{${\rm \not\! E}$}}
\newcommand{\met}{\mbox{${\rm \not\! E}_{\rm T}$}}

\title{Fake Dark Matter at Colliders}
\author{Spencer Chang}
\affiliation{Department of Physics, University of California, 1 Shields Avenue, Davis, CA 95616\\ and
Center for Cosmology and Particle Physics, Department of Physics, New York University, New York, NY 10003}
\author{Andr\'e de Gouv\^ea}
\affiliation{Northwestern University, Department of Physics \& Astronomy, 2145 Sheridan Road, Evanston, IL 60208-3112}
\preprint{NUHEP-TH/09-04}
\date{\today}
\begin{abstract}
If the dark matter (DM) consists of a weakly interacting massive particle (WIMP), it can be produced and studied at future collider experiments like those at the LHC. The production of collider-stable WIMPs is characterized by hard scattering events with large missing transverse energy. Here we point out that the discovery of this well-characterized DM signal may turn out to be a red herring. We explore an alternative  explanation -- fake dark matter -- where the only sources of missing transverse energy are standard model neutrinos.  We present examples of such models, focusing on supersymmetric models with $R$-parity violation.  We also briefly discuss means of  differentiating fake dark matter from the production of new collider-stable particles.   
\end{abstract}

\maketitle

\setcounter{equation}{0} \setcounter{footnote}{0}
\section{Introduction}
\label{sec:Intro}

One of the most important recent developments in fundamental physics is the confirmation that most of the matter in our universe is `dark.' Furthermore, this dark matter (DM) is made up of something other than the fundamental fields  that constitute the standard model of particle physics (SM) \cite{Amsler:2008zzb}. So far, however, all confirmed evidence for DM is astrophysical or cosmological, and all speak to its long-range gravitational effects. As far as current data are concerned, DM need not interact via SM interactions at all. If, however, the DM consists of a thermal relic, it is well-known that a weakly interacting electroweak-scale massive particle (WIMP) fits the bill quite well. In this case, there is the possibility of observing non-gravitational effects of DM. So-called direct  
and indirect 
DM searches have matured significantly over the past several years and are now sensitive to WIMPs with properties that would be characteristic of the DM (see, for example, \cite{Bertone:2004pz}). Indeed, both types of probes have recently revealed tantalizing results that may prove to be the first non-gravitational evidence for DM \cite{Bernabei:2008yi,Adriani:2008zr,:2008zzr}. 

On a different front, the Large Hadron Collider (LHC) will start producing -- very soon -- proton--proton collisions with unprecedented center-of-mass energies and luminosities and there is the possibility that WIMP DM particles will be directly produced. If one is able to detect WIMPs at the LHC detectors, it is widely believed that the combination of data from direct/indirect DM searches and the LHC will resolve the DM puzzle and revolutionize our understanding of fundamental physics. 

Many new physics theories beyond the standard model (BSMs) advocate the existence of new electroweak-mass degrees of freedom, can easily accommodate WIMP DM, and allow one to compute its relic density.  Astrophysical and cosmological measurements of the dark matter abundance constrain $\Omega_{\rm DM}=0.233\pm 0.013$ \cite{Hinshaw:2008kr} and provide nontrivial information that helps drive many of the phenomenological studies at the  LHC. Indeed, many of the discussions of new particle production and detection at the LHC assume the presence of a collider-stable WIMP within some more complex BSM.   

In a typical BSM, the LHC directly produces new color/charged states that ultimately decay into light SM particles and the WIMP. In this case, characteristic events consist of leptons and/or jets plus a sizable amount of missing transverse energy $E_{T}^{\rm miss}$. $E_{T}^{\rm miss}$ is interpreted as the production of a WIMP that is stable and exits the detector unscathed.  Given the WIMP motivation, it will be very tempting to label non-SM-like events with large $E_{T}^{\rm miss}$ as candidates for DM production or, at least, due to new collider-stable states.  Many studies have been done to determine the BSM from these LHC events (see for e.g. \cite{ArkaniHamed:2005px,Hubisz:2008gg}) and  ascertain the relic density \cite{Baltz:2006fm}.  

The observation of large $E_{T}^{\rm miss}$ at colliders, however, need not be related to DM or even new collider-stable particles. What appears to be a {\it bona fide} ``DM collider signal'' may, in fact, turn out to be a red herring.  As a conservative alternative, the missing energy may, instead, be due to SM neutrinos.  Any new physics model that leads to the nonstandard production of neutrinos can potentially fake the DM signal.  We call such new physics scenarios `fake dark matter' (FDM).  

Here, we discuss BSMs where final state neutrinos lead to  $E_{T}^{\rm miss}$ signatures that mimic well-characterized WIMP signatures, and attempt to identify ways of distinguishing the two hypotheses. In Sec.~\ref{sec:definition}, we properly define, in a model-independent way, what we mean by fake dark matter. In Sec.~\ref{sec:models}, we explore concrete examples and point out that many of the best known BSMs can, in fact, lead to fake dark matter events at colliders. We concentrate our discussion on supersymmetric extensions of the SM. In Sec.~\ref{sec:pheno}, we describe the collider prospects for disentangling a real DM signal from a fake one. Among other clues, we briefly discuss the possibility  of constraining the mass of the particle responsible for $E_{T}^{\rm miss}$. In Sec.~\ref{sec:conclusion} we offer some final remarks.

\setcounter{equation}{0}
\section{Fake Dark Matter}
\label{sec:definition}

\begin{figure}
\begin{center}
\includegraphics[width=8.5cm]{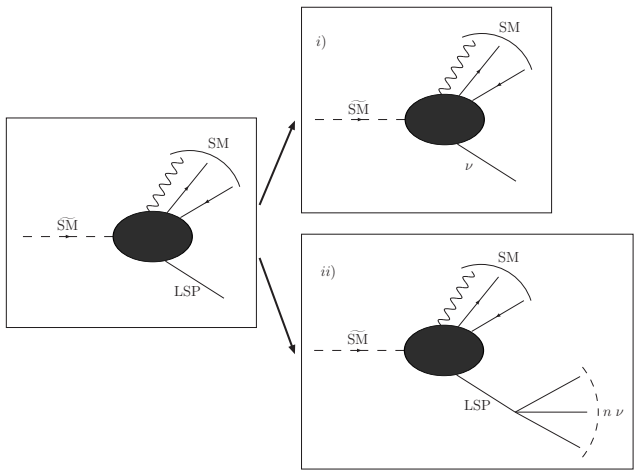}
\end{center}
\caption{Different decay modes of new, charged/colored heavy degrees of freedom $\widetilde{\rm SM}$. The left panel depicts the ``standard'' BSM scenario with a WIMP candidate where $\widetilde{\rm SM}\to \rm SM + LSP$ (LSP is a new collider-stable particle). The right panels depict different manifestations of fake dark matter. In $i)$ the role of the LSP is played by a neutrino while in $ii)$ the LSP is unstable, decaying into $n$ neutrinos.  \label{fig:cascade}}
\end{figure}

The traditional WIMP signature we will attempt to mimic, with fake dark matter (FDM), is depicted in Fig.~\ref{fig:cascade}(left): a new colored/charged degree of freedom ($\widetilde{\rm SM}$) is produced at a collider, promptly decaying into standard model degrees of freedom (denoted by SM) plus a collider stable particle (LSP).\footnote{We warn readers that given its familiarity and our inability to avoid it, we will often use SUSY lingo and analogies in our discussions, even when the discussion does not specifically apply to a supersymmetric extension of the SM.}  Since most BSMs ensure the stability of DM via a parity-like symmetry ({\it e.g.}, $R$-parity), such new states are often pair-produced, yielding two such cascades in every event.

The FDM scenarios we wish to study fall into two distinct classes. In the first class, depicted in Fig.~\ref{fig:cascade}(top-right), the phenomenology is very similar to the WIMP cascade-case, but the LSP is replaced by a neutrino.  Thus the neutrino plays the role of what looks like the true LSP.  In the second class, depicted in Fig.~\ref{fig:cascade}(bottom-right), there is a potential LSP candidate that is unstable and decays into a number of neutrinos within collider time scales.  The first class is more prevalent in the theoretical literature, and provides the most handles as far as disentangling fake from real dark matter.  On the other hand, the second class seems to be rarer but is potentially more difficult to debunk.  

We will restrict our discussion to experimentally ``tricky'' manifestations of FDM where the event sample with $E_T^{\rm miss}$ cannot be easily identified with neutrino production. For this purpose, we define two FDM requirements. Our first requirement for FDM is that the new physics should not lead to too many events with little or no missing energy. If the new physics leads to a large sample with little or no missing energy, we assume that it will be rendered distinct enough for experimental analyses to associate the $E_T^{\rm miss}$ to neutrinos, either through reconstructing mass peaks or event counting.  This eliminates, for example, scenarios where new particles cascade-decay to SM particles through $W$ or $Z$-bosons and the source of large $E_T^{\rm miss}$ are neutrinos from $W/Z$-boson decays.  Such scenarios can be identified experimentally by comparing the relative size of different event samples with the hypothesis that zero, one, and two electroweak gauge bosons have decayed into neutrinos. The constraints due to this requirement on FDM models will become clear in the next section. 

The other of our FDM requirements is the absence of displaced vertices.  The presence of displaced vertices often makes event reconstruction easier and will reveal that one is not dealing with a characteristic collider-stable WIMP signature.  For example, displaced vertices are commonly associated with super weakly interacting massive particles that are not thermal relics \cite{Feng:2003xh}.  We will also comment on this requirement in the upcoming section.

\setcounter{equation}{0} 
\section{Examples of Fake Dark Matter} \label{sec:models}

Here we discuss examples of scenarios that may lead to an FDM signal. 

\subsection{WIMPless New Physics -- Leptoquark FDM \label{sec:newphysics}}
Before discussing complete BSMs, it is instructive to present a simple FDM scenario. This can be accomplished by adding to the SM a new heavy charged degree of freedom whose decay into SM particles always contains neutrinos. 

We will consider one scalar $SU(2)_L$-doublet leptoquark {\cite{Amsler:2008zzb}, $X_d$,\footnote{A very similar picture can be drawn with a $(3,2)_{+7/6}$ leptoquark $X_u$ which couples to SM fields via $\lambda_uX_uu^cL$, where $u^c$ is the up-type antiquark singlet field. In this case, we would also need to impose that the coupling associated to the interaction $X^*_u e^c Q$, where  $Q$ and $e^c$ are the left-chiral $SU(2)_L$ quark doublet and antilepton singlet respectively, is significantly smaller than $\lambda_u$ in order to construct an FDM scenario with few events with no missing energy.} which couples to standard model fermions via
\begin{equation}
\lambda_d X_d d^c L. 
\label{eqn:leptoquark}
\end{equation}
Here $L=\left(\ell,\nu\right)^T$, $d^c$ are the left-chiral $SU(2)_L$ lepton doublet and down-type antiquark singlet, respectively. $\lambda_{d}$ is a dimensionless coupling and generation indices have been omitted. Gauge invariance dictates that, under $SU(3)_c \times SU(2)_L \times U(1)_Y$, $X_d$ transforms as $(3,2)_{+1/6}$, which, after electroweak symmetry breaking, describes two colored states $x^{+2/3}_d$ and $x^{-1/3}_d$ with charges $+2/3$ and $-1/3$, respectively.

If $X_{d}$ decays are to qualify as FDM, other constraints need to be satisfied. Electroweak symmetry breaking will distinguish the up and down components of $X_{d}$ and split their masses-squared:
\begin{eqnarray}
&m_{-1/3}^2=M_d^2, \\
&m^2_{+2/3}=M_d^2+ \Delta v^2, 
\end{eqnarray}
where $M_d^2$ is the common mass-squared for the two states, $\Delta$ is a dimensionless coupling related to the scalar potential and $v$ is the vacuum expectation value of the neutral component of the SM Higgs doublet.  

The interactions contained in Eq.~(\ref{eqn:leptoquark}) allow $x^{+2/3}_d$ to decay into a quark and a charged lepton.  In order to construct an FDM scenario and hence suppress this visible decay, we first choose $\Delta>0$ rendering $x^{-1/3}_d$ lighter than  $x^{+2/3}_d$ and allowing $x^{+2/3}_d$ to  weak-decay into the $x^{-1/3}_d$ state and a $W$-boson. FDM requirements point to the region of the parameter space (scalar masses and the coupling $\lambda_d$) where the weak decay dominates over the $\lambda_d$ mediated one. Failed searches for leptoquarks constrain the masses of the $x_d$'s while electroweak precision data constrain their mass-squared difference. In particular, contributions to the $T$-parameter are proportional to $m_{+2/3}^2-m_{+1/3}^2$. For $m_{+2/3}=400$~GeV, we estimate that $\delta T<0.1$  translates roughly into $\Delta < 0.5$ (or $m_{-1/3}>355$~GeV).  For experimentally allowed values of $x^{+2/3}_d$, $x^{-1/3}_d$ masses, the weak decay of the $x^{+2/3}_d$ to  $x_d^{-1/3}$ will proceed through an offshell $W$-boson, resulting in a three-body final state. 

In summary, experimental constraints plus $\Delta>0$ imply that the leptoquarks $x_d$ decay as follows: 
\begin{eqnarray} 
& x^{-1/3}_d & \to d+\bar{\nu}, \\
& x^{+2/3}_d & \to d+\ell^+,  \\
& & \to W^{+*}+x_d^{-1/3},
\end{eqnarray}
where $d$ is a down-type quark, $\ell$ is a charged lepton and $W^{+*}$ indicates that the $W$-boson is offshell. The requirement that $x^{+2/3}_d$ decays lead to final-state neutrinos most of the time, $B(x^{+2/3}_d\to d+\ell)\ll B(x^{+2/3}_d\to x_d^{-1/3}+W^*)$, implies
\begin{eqnarray}
\lambda_d^2\ll & \frac{9\times 8}{15\pi^2}\frac{(m_{+2/3}-m_{-1/3})^5}{v^4m_{+2/3}}, \nonumber \\ \ll & 8\times 10^{-9}\left(\frac{\Delta}{0.1}\right)^5\left(\frac{400~\rm GeV}{m_{+2/3}}\right)^6,
\label{3-body}
\end{eqnarray}
where a factor of 9 accounts, very roughly, for all different allowed ``decays'' of the offshell $W$-boson and we made the assumption that $m_W^2\gg m^2_{+2/3}-m^2_{-1/3}$ (or $\Delta \ll 0.1$).\footnote{While the simple expression Eq.~(\ref{3-body}) is only valid for small $\Delta$ values, it should be clear that even for $\Delta\sim 1$ all weak $x_d^{+2/3}$ decays involve offshell $W$-bosons. In these cases, there are points in the parameter space where $x_d^{+2/3}$ decays virtually all the time into $x_d^{-1/3}$, while the $x_d^{-1/3}$ decay is prompt.}  We further wish the $x_d^{-1/3}$ to decay without a displaced vertex. Typical impact parameter resolutions are of order $10\, \mu$m so we require $\lambda_d^2/(8\pi)m_{-1/3}\gg 1/(10\, \mu {\rm m})\sim 2\times 10^{-11}$~GeV or  
\begin{equation}
\lambda_d^2\gg 10^{-12}\left(\frac{250~{\rm GeV}}{m_{-1/3}}\right). 
\label{lambda-}
\end{equation}
There are regions in the new physics parameter space where Eqs.~(\ref{3-body}) and (\ref{lambda-}) {\sl can} be satisfied simultaneously.

We conclude that, unless $\Delta$ is very small, one can find $\lambda_d$ such that the lightest $X_d$ particle decays promptly into a jet plus a neutrino, while the heaviest $X_d$ particle decays mostly into the lightest $X_d$ particle plus an offshell $W$-boson.  Furthermore, as discussed above, electroweak precision constraints can be easily evaded. The FDM constraints on  $\lambda_{d}$ discussed above render it small enough that one satisfies results from searches for charged lepton flavor violation (especially $\mu\to e$-conversion in nuclei if the associated quark fields are of the first generation), searches at HERA, and other experimental constraints. Current constraints are summarized in \cite{Amsler:2008zzb}.

Under these circumstances, $x_d$'s will be strongly produced in pairs at the LHC. 
After production, the $x_d$ particles will decay in one of two different ways. The lightest $x_d$ particle decays into a hard jet plus missing energy (leading to $p+p\to 2~{\rm jets}+E_T^{\rm miss}$). The heaviest $x_d$ particle ``beta-decays'' into its isospin counterpart plus two light fermions. This leads to $p+p\to 2~{\rm jets}+W^{-*}+W^{+*}+E_T^{\rm miss}$, where $W^*$ stands for the decay products of an offshell $W$-boson, which most of the time is hadronic. Overall, $x_d$ production is rather similar to squark pair production at the LHC, followed by $\tilde{q}\to q+{\rm LSP}$ or $\tilde{q}\to q'+\chi^{\pm}\to q'+W^{\pm(*)}+{\rm LSP}$, where $\chi^{\pm}$ are charginos and LSP is the lightest supersymmetric particle. In the latter case, there is an interesting difference that could be exploited in order to distinguish the fake from the genuine dark matter scenario. In the case of SUSY, the $W$-boson and the invisible particle form a resonance (assuming the chargino is onshell), while in the leptoquark case the invisible particle and the jet form a resonance. 

Other experimental probes may be able to distinguish whether $x_d^{-1/3}$ is decaying into a stable heavy particle or a neutrino. A high energy $e^+e^-$ collider should see $e^+e^-\to$~jets mediated by $t$-channel $x_d$ exchange if $x_d$ couples to first generation leptons, while neutrino deep-inelastic scattering should also be sensitive to $\nu+q\to \nu+q$ mediated by (offshell) $x_d$ exchange. In this case the flavor of the final state quark or neutrino need not agree with that of the target quark or beam neutrino. For a recent study of a next-generation, high energy neutrino scattering experiment, see \cite{Adams:2008cm}.

\subsection{Supersymmetric FDM -- $R$-parity Violation}
The MSSM with $R$-parity violation (RPV) has been extensively studied for its collider signatures.  For a recent comprehensive review see, for example, \cite{Barbier:2004ez}. For the most part, the phenomenology is expected to be quite distinct from the MSSM without RPV, but there are exceptions. For example, if the MSSM superpotential contains $LLE$ terms ({\it cf.} Eq.~(\ref{rpv_super})) and the lightest neutralino is the lightest superpartner (LSP), each SUSY cascade is guaranteed to contain some amount of $E_{T}^{\rm miss}$ in the form of at least one neutrino.  A distinctive property of this scenario is that each decay is also characterized by a potentially large number of charged leptons \cite{Barbier:2004ez}.  Other previously unknown -- to the best of our knowledge -- regions of RPV space can mimic $R$-parity conserving SUSY and qualify as FDM. These will be discussed momentarily.  

We will consider the consequences of ``turning on'' one RPV interaction at a time, and concentrate on those that contain neutrino superfields.  We will focus exclusively on renormalizable RPV couplings in the superpotential.  Kahler couplings should also be explored, but since chiral suppression tends to reduce couplings to neutrinos these will be ignored henceforth. As in the previous subsection, we will identify conditions for the different RPV couplings so that the collider signals qualify as FDM. We will first consider only the MSSM particle content, where all neutrino fields reside inside the lepton-doublet chiral superfields $L$. We then explore scenarios with small neutrino masses where RPV interactions containing gauge-singlet chiral superfields $N$ are present. 

\subsubsection{MSSM}
Given the MSSM particle content,  the renormalizable RPV superpotential interactions involving neutrino fields  include\footnote{Bilinear RPV, $\mu' L H_u$, will not be considered. Blinear RPV FDM scenarios are identical to those with trilinear RPV. The reason is that one can perform a superfield redefinition where $\mu'$ is chosen zero while $\lambda$ and $\lambda'$ terms are not \cite{Hall:1983id}. 
SUSY breaking effects, which prevent one from perfectly mapping the bilinear RPV scenario into the trilinear ones, do not lead to new manifestations of FDM.}  
\begin{equation}
\lambda_{ijk}\; L^i L^j E^{c,k} + \lambda'_{ijk}\; L^i Q^j D^{c,k} \label{rpv_super},
\end{equation}
where $\lambda,\lambda'$ are dimensionless couplings, $i,j,k$ are flavor indices and $Q,L,D^c,E^c$ are chiral superfields associated to the left-handed quarks, leptons, down antiquarks and charged antileptons.
We proceed to identify regions of the parameter space that lead to FDM. In most cases the RPV couplings $\lambda$ or $\lambda'$ are constrained to be small enough that all MSSM production and decay processes are dominated by the $\lambda=\lambda'=0$ Lagrangian until one reaches the LSP. Hence, often the only impact of the RPV coupling is to govern the decay of the LSP. We discuss the phenomenology with the two types of RPV couplings turned on separately.  

\underline{$LQD$} -- In terms of component fields, the $LQD^c$ term in the superpotential contains
\begin{equation}
{\cal L}_{LQD}\supset \lambda' \left[\left(\nu d-\ell u\right)\tilde{d^c}+\left(\nu \tilde{d}-\ell \tilde{u}\right)d^c+\left(\tilde{\nu} d-\tilde{\ell} u\right)d^c\right].
\end{equation}
Allowed RPV sparticle decay vertices are depicted in Fig.~\ref{fig:LQD}.  On the left are partially invisible decays, containing a neutrino, while on the right are the visible decays without neutrinos.   
\begin{figure}
\begin{center}
\includegraphics[width=6.5cm]{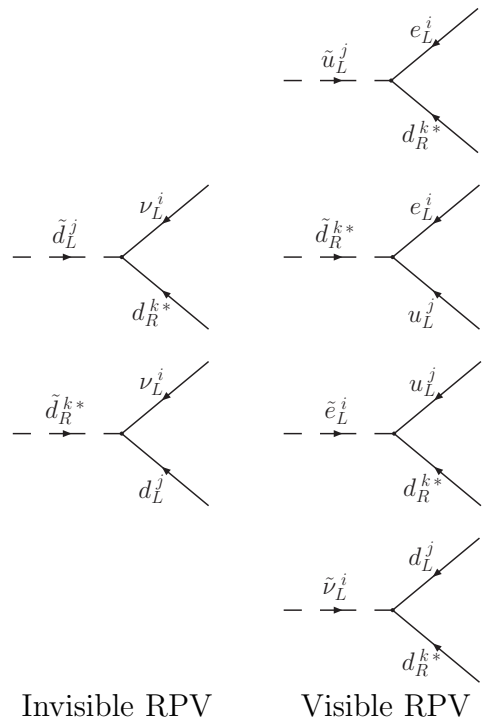}
\\{\large  Invisible RPV \hspace{0.9cm} Visible RPV} 
\end{center}
\caption{$LQD^c$ induced decays, with partially invisible decays on the left and visible decays on the right. \label{fig:LQD}}
\end{figure}

Since we are interested in LSP decays that lead to neutrinos in the final states, we first consider the case where the LSP candidate is a right-handed or left-handed down-squark. The case of left-handed down squarks is similar to the $X_d$ case discussed in Sec.~\ref{sec:newphysics}, with a few important differences. In the MSSM case it is easy to see that, ignoring flavor and left-right mixing effects, only third generation down-type squarks $\tilde{b}_L$ can be lighter than its up-type squark partner $\tilde{t}_L$ (for $\tan\beta>1$), and their mass-squared splitting is of order $m_t^2$. If there is no scalar top mixing, this would mean that all scalar tops heavier than 220~GeV  would decay either by visible RPV interaction or via charged-current weak interactions into a three-body final state.  As discussed in Sec.~\ref{sec:newphysics}, the offshell $W$-decay can dominate over the RPV-mediated decay as long as $\lambda'$ is small enough. We conclude that there is an allowed region of the RPV parameter space where all the FDM conditions are met. 

The presence of scalar top mixing, which is expected if the soft SUSY-breaking scalar masses-squared are not much larger than the top quark mass squared, requires some extra care, but does not spoil the possibility that a $\tilde{b}_L$ LSP leads to FDM in the presence of RPV. In all cases, however, we are constrained to soft-SUSY breaking third generation left-handed squark masses-squared that are not much larger than the top quark mass. 

At the LHC, a typical SUSY event would look like $p+p\to \tilde{b}\tilde{b}^*$ followed by $\tilde{b}\to\nu+$~jet.  If the $LQD$ coupling is flavor-aligned between $Q$ and $D$, the jet will be a bottom jet but, in general, the final state jet need not contain heavy flavors. Another possibility includes $p+p\to \tilde{t}\tilde{t}^*$ followed by $\tilde{t}\to\tilde{b}+W^*$. The offshell $W$-boson manifests itself as either two jets or an $\ell+\nu$ pair, with $\tilde{b}\to\nu+$~jet providing the missing energy signature.

This phenomenology mimics some scenarios with low-scale mediation of SUSY breaking like gauge-mediated SUSY breaking (GMSB).  In such models, the gravitino is the LSP and is weakly coupled to the MSSM.  These conditions allow the standard superpartner production and cascade decays to occur until the next-to-lightest supersymmetric particle (NLSP) is produced. The NLSP then decays to its SM partner and the gravitino; this associated partner acts as a particle tag of the decay.  However, such decays are often displaced and potentially outside the detector, since
\begin{equation}
c\tau_{\rm NLSP} \sim 100\, \mu {\rm m} \left(\frac{100\ \rm GeV}{m_{\rm NLSP}}\right)^5 \left(\frac{m_{3/2}}{2.4\ \rm eV}\right)^2
\end{equation}
where $m_{3/2}$ is the gravitino mass.

The FDM scenario described above is similar. If $\lambda'$ is small enough, SUSY particle production and cascade decays proceed ``normally'' until LSP's ($\tilde{b}_L$) are produced. These then decay through the RPV coupling into a neutrino plus a SM particle. However, this decay can be prompt and a visible particle tag may or may not be present.  Bottom jets can be produced in association with the sbottom LSP, but the sbottom decays do not necessarily yield bottom jets, unless the coupling is flavor aligned. This feature may provide a means of telling low-scale SUSY breaking from RPV $\lambda'$ FDM. Another interesting feature worth exploring is the fact that in gauge mediation right-handed sfermion masses are lighter than left-handed ones, while in the FDM scenario spelled out above we have the left-handed sbottom lighter than the right-handed one. $LQD$ RPV interactions are also expected to show up in other particle physics experiments, including neutrino deep inelastic scattering and searches for lepton-number violation. However, the potential absence of first generation quarks in $LQD$ would significantly inhibit signals at most of these experiments. 

There is also the possibility that a right-handed down-type squark is the LSP. In order to realize the FDM scenario, one needs to suppress the allowed $\tilde{d}_R\to e u$ visible decay mode to avoid events without $E_T^{\rm miss}$. The only way to accomplish this is to again assume that the RPV coupling involves the top quark in $Q$. This way, $\tilde{d}_R\to e t$ may be kinematically suppressed with respect to $\tilde{d}_R\to \nu b$ as long as the $\tilde{d}_R$ is light enough. As an example, if the partially invisible decay width (the one associated to neutrinos in the final state) is to be larger than 90\% of the total width, we need $m_{\tilde{b}_R}/m_t<1.06$.  In this case, we do get the particle tag of the bottom jet, even though the LSP may not be a bottom squark.   

We will briefly comment on some other scenarios where the LSP is not a down-type squark. The FDM gluino LSP decay is
\begin{equation}
\tilde{g}\to \bar{b}+d+\nu,
\label{gluino_LSP}
\end{equation}
where we assume that the $LQD$ coupling involves only third generation left-handed squarks. The unwanted visible decay $\tilde{g}\to \bar{t}+d+e$ can be suppressed either by phase space (gluino not much heavier than the top-quark) or by postulating that the left-handed sbottom is much lighter than the left-handed stop. Also, the unwanted visible decay $\tilde{g}\to \bar{b}+e+t$ mediated by the right-handed down squark can be suppressed by either making the right-handed down squark much heavier than the left-handed sbottom or again by kinematics. Qualitatively similar scenarios can be realized in the case of $\chi^+$ or $\chi^0$ LSP. In all cases one needs to make sure that the LSP decay is prompt. Finally, if a right-handed slepton were the LSP, $LQD$-mediated decays would lead to four fermions in the final state (we are assuming that left-right slepton mixing is negligibly small and that left-handed sleptons are much heavier than the squarks). In this case, it appears very unlikely that the LSP decays are prompt. 

\underline{$LLE$} -- In terms of component fields, the $LLE^c$ term in the superpotential contains
\begin{equation}
{\cal L}_{LLE}\supset \lambda \left[\left(\nu \ell'-\nu' \ell\right)\tilde{e^c}+\left(\tilde{\nu} \ell'-\nu' \tilde{\ell}\right)e^c+\left(\nu \tilde{\ell}'-\tilde{\nu}' \ell\right)e^c\right].
\end{equation}
Allowed RPV sparticle decay vertices are depicted in Fig.~\ref{fig:LLE}. Notice that the flavor of $\ell,\nu$ must differ from that of $\ell',\nu'$.
\begin{figure}
\begin{center}
\includegraphics[width=7cm]{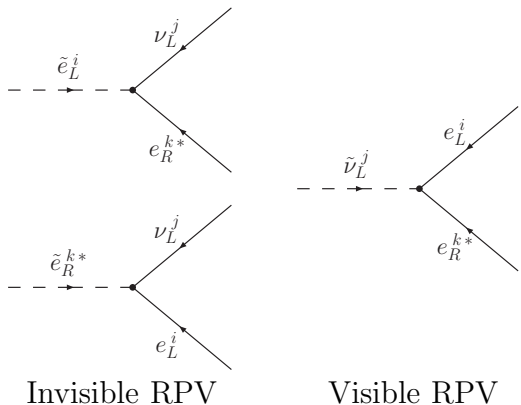}
{\large \hspace{.2cm} Invisible RPV \hspace{1.2cm} Visible RPV} 
\end{center}
\caption{$LLE^c$ induced decays, with invisible decays on the left and visible decays on the right. The decays where $i$ and $j$ are reversed are not shown. \label{fig:LLE}}
\end{figure}

Unlike the $LQD$ case, several FDM scenarios seem readily apparent. Here we concentrate on the case where right-handed charged sleptons are much lighter than left-handed ones. This way, RPV mediated LSP decays are guaranteed to contain neutrinos in the final state as long as the coupling of the LSP to right-handed sleptons is not too small. 

Choices for the LSP are many, and the FDM phenomenology will differ for different LSPs.  The most obvious choice for the LSP is the $\tilde{e}_R$, where $e$ stands for the selectron, the smuon or the stau. The phenomenology in this case is identical to low-scale GMSB scenarios with staus as the NLSP, except that the lepton flavors do not need to ``match.'' For example, ignoring flavor violating effects, a stau NLSP in the case of GMSB decays almost exclusively into tau leptons and a gravitino. In this stau LSP RPV scenario, the staus decay into electrons, muons and taus (plus different flavored neutrinos) with potentially unrelated branching ratios. Bounds on the $\lambda$ couplings (from neutrino masses, charged-lepton flavor-violation, neutrino scattering, etc) do not preclude large branching ratios into all lepton flavors as long as all $\lambda$ couplings are smaller than $10^{-3.5}$ or so \cite{Barbier:2004ez}. 

Another scenario, already discussed in the literature \cite{Barbier:2004ez},  is the lightest neutralino $\tilde{\chi}_1^0$ as the LSP. In this case, $\chi_1^0\to \ell\ell'\nu''$ is the only allowed LSP decay mode, and the relative branching ratios to different flavor final states will depend on the relative sizes of the different RPV $\lambda$ couplings. Given the three-body-final-state nature of the LSP decay, one needs to worry about how prompt its decay is. Roughly,
\begin{equation}
\Gamma_{\rm LSP}\sim \frac{\lambda^2g'^2}{100\pi^3}\frac{m^5_{\tilde{\chi}_1^0}}{m^4_{\tilde{e}_R}}\sim 10^{-4} \lambda^2 \left(\frac{m_{\tilde{\chi}_1^0}}{250~{\rm GeV}}\right)~{\rm GeV},
\end{equation} 
where we set $m_{\tilde{e}_R}=2m_{\tilde{\chi}_1^0}$. $\Gamma\gg 1/(10\, \mu {\rm m})$ implies $\lambda^2\gg 10^{-7}$ for SUSY particle masses in the few to several hundred GeV range. This is currently allowed by experimental data, but is already constrained for certain combinations of $\lambda$ couplings. Note that one would be faced with a similar scenario in the case of a chargino LSP.  

Scalar quark and gluino LSP's are trickier. Squarks decay into a four-body final state via $\tilde{q}\to q\tilde{\chi}^*\to q(\ell\ell'\nu'')$ while gluinos decay into a five-body final state. Given current bounds on RPV couplings, it is very unlikely that either of these LSPs would decay promptly.  Many of these cases where $\tilde{e}_R$ is not the LSP share phenomenological signatures with scenarios where the right-handed sneutrino is the LSP (for recent discussions see \cite{deGouvea:2006wd,Gupta:2007ui,Choudhury:2008gb}). 

It is worth noting that some states do not couple to right-handed scalar leptons (say, pure winos). Once produced, these would decay via (potentially offshell) left-handed sleptons and ultimately to either $\ell\ell'\nu''$ or $\ell\ell'\ell''$. The fraction of the $E_T^{\rm miss}=0$ new physics events would depend on the relative masses of left and right handed sleptons and the size of the RPV coupling, similar to the $\tilde{b}_L$ and $X_d$ cases discussed earlier. 

Regardless of the nature of the LSP,  the MSSM augmented by $LLE$ RPV couplings will lead to an abnormal amount of events with charged leptons in the final state.  Thus, this type of FDM would be disfavored if one were to also encounter a large sample of jets plus large $E_T^{\rm miss}$ and no charged leptons.

\subsubsection{MSSM with Right Handed Neutrinos}

With the addition of new degrees of freedom to the MSSM field content, other FDM scenarios materialize. A simple and very well motivated extension of the MSSM is the addition of singlet chiral superfields $N$. Gauge invariance allows the following renormalizable terms in the superpotential:
\begin{eqnarray}
& W_{N} = & fN+\frac{k}{3}N^3+\lambda_HNH_uH_d + \nonumber \\ && +y_{\nu}LH_uN+\frac{M_N}{2}N^2, 
\label{WN}
\end{eqnarray}
where flavor indices have been suppressed and $f,k,\lambda_N,y_{\nu},M_N$ are free parameters. Global symmetries dictate which among the couplings above are ``turned on.'' If the $N$ superfields are $R$-odd, the first line of Eq.~(\ref{WN}) above is absent and the $N$'s are referred to as right-handed neutrino superfields. In this scenario, neutrinos acquire non-zero masses, as experimentally required, either of the Majorana  kind ($M_N\neq0$) or the Dirac kind ($M_N=0$). If the $N$ superfields are $R$-even, all $y_{\nu}$ are forbidden. The imposition of a $Z_3$ symmetry would lead to $f=M_N=0$ and a vanishing $\mu$-term in the MSSM superpotential. In this case we are left with the well-known next-to-minimal supersymmetric standard model (NMSSM) \cite{Ellis:1988er}. The NMSSM provides a dynamical mechanism for generating the $\mu$-term during electroweak symmetry breaking and leads to several interesting phenomenological consequences. Here we are interested in the former case, where the $N$'s ``look like'' right-handed neutrino superfields and where the RPV couplings $\lambda_N$ and $k$ are considered small.\footnote{Henceforth, we will not discuss the impact of the tadpole terms $fN$.} Detailed discussions of different aspects of Eq.~(\ref{WN}) -- mostly concentrating on the spectrum of fermions, sfermions and Higgses -- and variations thereof can be found in \cite{Kitano:1999qb,LopezFogliani:2005yw,Chemtob:2006ur,Mukhopadhyaya:2006is,Escudero:2008jg,Ghosh:2008yh}.

Unlike the MSSM cases discussed earlier, identifying FDM scenarios once $N$ superfields are added to the Lagrangian is very non-trivial. The phenomenology of these scenarios depends significantly on the details of how neutrino masses are generated (governed in part by the magnitude of  $M_N$ and $y_{\nu}$), the magnitude of left-right sneutrino mixing (which depends on the magnitude of the SUSY-breaking $A$-terms), and the magnitude of the RPV effect. It is important to note that we must require $M_N$ of order or smaller than the weak scale. Otherwise, both right-handed neutrinos and sneutrinos acquire masses of order $M_N$ and decouple from the theory at the weak scale, leaving behind non-renormalizable RPV interactions which are typically too weak to realize FDM scenarios.

As in the MSSM case, we managed to find different FDM scenarios for different LSP candidates by considering only a single RPV coupling at a time.   
Due to the additional layers of complexity associated to the $N$ superfields and the generation of neutrino masses,  the situation here is described more qualitatively than the MSSM cases of the previous subsection.

\underline{$NH_uH_d$} --  The $\lambda_N$ superpotential term contains the following couplings:
\begin{eqnarray}
& {\cal L}_{NH_uH_d} & \supset \lambda_N \left[\left(\tilde{h}_u^+\tilde{h}_d^--\tilde{h}_u^0\tilde{h}_d^0\right)\tilde{\nu}^c \right. \label{nhh}\\
&& \left. +\left(h_u^+\tilde{h}_d^--h_u^0\tilde{h}_d^0 +\tilde{h}_u^+h_d^--\tilde{h}_u^0h_d^0\right)\nu^c
\right], \nonumber  
\end{eqnarray} 
where $\nu^c$ is the right-handed neutrino field, $\tilde{\nu}^c$ the right-handed sneutrino field, $\tilde{h}_{u,d}$ are the different Higgsino fields, and $h_{u,d}$ are the Higgs fields. After electroweak symmetry breaking, right-handed neutrinos mix with left-handed neutrinos and the neutral Higgsinos, acquiring Majorana masses regardless of whether $M_N$ is non-zero since the $\lambda_N$ term violates lepton number. If the $\tilde{\nu}^c$ acquire vacuum expectation values, charged leptons also mix with the charged Higgsinos. Throughout this discussion we will assume, for simplicity, that mostly active neutrino masses are generated, predominantly, by the canonical seesaw contribution, {\it i.e.}, $m_{\nu}\sim y_{\nu}^2v^2/M_N$. 

We briefly consider one scenario. If the lightest neutralino is the LSP and if sneutrino--Higgsino mixing is small, one expects to observe the standard MSSM decay chains of sparticles into SM particles plus the LSP, $\chi^0_1$. Neutralinos then decay predominantly via $\chi^0_1\to \nu h^0$ and $\chi^0_1\to \nu Z^0$, mediated by the second line in Eq.~(\ref{nhh}). $h^0$ refers to either one of the neutral scalar Higgs bosons and the $Z$-boson decay is induced via neutrino--Higgsino mixing. If right-handed neutrinos are not kinematically accessible, this decay is suppressed by the mixing angle between left-handed and right-handed neutrinos, which is, naively, proportional to $\sqrt{m_{\nu}/M_N}$, where $m_{\nu}\lesssim 0.1$~eV are the mostly active neutrino masses. If $M_N> 100$~GeV, we estimate $c\tau\gtrsim(10/\lambda_N^2)~\mu$m  for $m_{\chi_1^0}\sim 100$~GeV. This means that prompt $\chi^0_1$ decays are only possible if it can decay into an onshell $\nu_R$\footnote{For very small or vanishing $M_N$, neutrino mass generation is potentially more complicated. In this case, Higgsinos and right-handed neutrinos mix (with a Majorana mass proportional to $\lambda_Nv$) along with the left--right neutrino mixing induced by the Yukawa interaction (Dirac mass proportional to  $y_{\nu}v$).} and if $\lambda_N$ values are not too small. 

We are therefore confined to a low-scale seesaw, and the dominant LSP decay is $\chi^0_1\to \nu_R (h^0,Z^0)$. It remains to check the decay of $\nu_R$. Assuming their decay is dominated by the left-right neutrino mixing contribution proportional to $\sqrt{m_{\nu}/M_N}$, 
\begin{equation}
c\tau_{\nu_R}\sim \left(\frac{10~\rm GeV}{M_N}\right)^4\times 1000~\rm m,
\end{equation}
which implies that $\nu_R$ are collider stable for $M_N\lesssim 40$~GeV.   If $\nu_R$ is light enough, $\lesssim$ few GeV, from the collider perspective it will behave like a standard neutrino (collider-stable, effectively massless, neutral, spin one-half, etc) and serves as FDM.  However, if its mass is heavy enough to be measurable at a collider, it will appear as a new massive state escaping the detector and thus will not be seen as FDM.   Such phenomenology is very similar to scenarios with a light gravitino and a Higgsino NLSP \cite{Matchev:1999ft}, where the Higgs and $Z$-bosons are produced in the Higgsino decay to the gravitino.       

If the mostly right-handed sneutrino was the LSP, it would decay in a variety of ways (through chargino and neutralino loops, or via lepton--Higgsino mixing) into gauge bosons (including photons), charged fermions and neutrinos. Therefore, right-handed sneutrino LSP decays do not always yield final states with large $E_T^{\rm miss}$ and do not realize FDM in the strict sense.  However, since sneutrinos at hadron colliders are often produced in association with neutrinos, most events will have additional missing energy.  The rest of the phenomenology of right-handed sneutrino LSP scenarios is similar to that of scenarios where $R$-parity is conserved and the right-handed sneutrino is the LSP.

\underline{$N$}$^3$ -- The $k$ superpotential term contains the following coupling:
\begin{equation}
{\cal L}_{N^3}\supset k\nu^c\nu'^c\tilde{\nu}''^c.
\end{equation}
In this case, if the mostly right-handed sneutrino is the LSP, it decays, potentially promptly, into right-handed neutrinos, assuming that those are kinematically accessible. For concreteness, we will concentrate on the possibility that the neutrinos are Dirac (or pseudo-Dirac) fermions. This can be realized if the right-handed sneutrinos do not acquire a vacuum expectation value and all $M_N$ vanish. In this case, while lepton-number is violated, a $Z_3$ symmetry, which may be exact depending on the details of SUSY breaking, forbids both $\nu_L$ and $\nu_R$ from acquiring Majorana masses.  

A right-handed sneutrino LSP in this case looks like the second class of FDM scenarios described in Sec.~\ref{sec:definition}. The rest of the phenomenology is identical to the MSSM with a predominantly right-handed sneutrino as the LSP and depends significantly on the size of left-right sneutrino mixing. If left-right sneutrino mixing is large, a typical SUSY signal at the LHC would be, say, squark production followed by $\tilde{q}\to\chi_1^0$+jet with $\chi_1^0\to\nu\tilde{\nu}$. Note that both the $\chi_1^0$ and the LSP $\tilde{\nu}$ decay invisibly and, if left-right sneutrino mixing is large, both decays are prompt. This scenario is, at a hadron collider, indistinguishable from a neutralino LSP and $k\neq0$. In this case, $\chi_1^0\to\bar{\nu}\nu\nu$ via an offshell sneutrino. The two possibilities (neutralino versus mostly right-handed sneutrino LSP) may be distinguishable with a linear collider if the left-right sneutrino mixing is large. In this case, along with lightest neutralino pairs ($e^+e^-\to \chi_1^0\chi_1^0$), both sneutrinos  can be produced at an $e^+e^-$ machine: $e^+e^-\to \tilde{\nu}_a+\tilde{\nu}^*_b$ where $a,b=R,L$ indicate the mostly left-handed and mostly right-handed sneutrinos. In principle, the linear collider should be able to reveal that there are two invisible states ($\tilde{\nu}_R$ and $\chi_1^0$) and establish whether the lightest one is $\chi_1^0$ (assuming that the mass of $\chi_1^0$ has been estimated at the LHC). The process $e^+e^-\to$~invisible, will not, of course, yield any information, but this can be mitigated by studying higher order processes like $e^+e^-\to\gamma+E_T^{\rm miss}$ or even $e^+e^-\to Z^0+E_T^{\rm miss}$.\footnote{The latter may also reveal whether the invisible state couples to the $Z$-boson.}

In order to test whether the LSP is decaying into neutrinos, one needs to ``see'' an effect mediated by the coupling $k$. This requires observing the neutrino coupling to the LSP. In principle, the sneutrino decay into neutrinos can be probed at a high energy  lepton collider through the following neutrino fusion process: $e^-e^-\to W^-+W^-+\tilde{\nu}_R$.  Here the electrons convert into $W$-bosons plus neutrinos which annihilate into the predominantly right-handed sneutrino. The sneutrino mass can be reconstructed by measuring the outgoing $W$-bosons and requiring conservation of energy-momentum. This process, alas, relies on mass insertions on both neutrino legs and is, hence, completely negligible.

\setcounter{equation}{0} 
\section{LHC Signals -- Can We Tell if Our Dark Matter is Fake?} \label{sec:pheno}

If the LHC sees a large sample of missing energy events that are unexplained by Standard Model sources, it will be important to determine whether the missing energy is due to new collider-stable particles.  As a step in this process, it should be determined if fake dark matter scenarios are consistent with the data.  Some model dependent tests are possible, as was illustrated in Sec.~\ref{sec:models}.  Many of the FDM SUSY scenarios predict specific particle tags in each new physics cascade decay, including charged leptons, $b$ jets, Higgs and $W, Z$-bosons.  Observing substantial missing energy events without such tags effectively rules out these particular scenarios. Once experiments have accumulated large data sets, more detailed analyses can be performed in order to test for different FDM scenarios.  For example, $LLE$ RPV scenarios can have pronounced asymmetries in the flavor distribution of the final state leptons which may help distinguish it from standard $R$-parity conserving SUSY.  

Given that FDM models are relatively unexplored, it is important to consider model-independent tests for neutrino production.  The most obvious method relies on the fact that neutrinos are effectively massless at colliders.  In cascade decays of the type depicted in Fig.~\ref{fig:cascade}(top-right), a single neutrino is produced at the end of each cascade decay.  Measuring the missing energy particle mass to be consistent with zero would suggest an FDM scenario of this type (and eliminate many dark matter models), while measuring the mass to be nonzero determines that only FDM models of the type depicted in Fig.~\ref{fig:cascade}(bottom-right) are allowed.

There is an extensive (and growing) literature on methods to reconstruct the mass of collider-stable particles.  As an early foray into possible analyses, we will discuss a few idealized analyses which, according to recent studies, may work.   While we will concentrate on a few scenarios within a particular model (SUSY with RPV), other models with the same kinematics (onshell versus offshell decays) can be analyzed with the same method.  To set the stage, we give a quick overview of the mass reconstruction techniques that have been proposed.  In processes involving many decays producing several visible particles, the endpoints of various invariant-mass distributions can be used to determine all the masses (see for e.g. \cite{Bachacou:1999zb}).  More recently, new methods were proposed. These have been shown to work with shorter cascades as long as two identical ones occur in each event.  These are the methods  on which we concentrate here. 

In the case where two or more visible particles are present in a cascade decay, one can use ``kinematic'' methods. Here one uses the observed visible particle momenta and the missing transverse momentum, while imposing invariant-mass constraints related to all onshell intermediate and final  states, to determine which mass values for the missing particles are consistent with the individual events \cite{Kawagoe:2004rz,Cheng:2007xv,Cheng:2008mg}.  Most studies have looked at situations where each decay of the cascade is two-body.  In the case of three visible particles per decay chain, one can directly solve for the masses by looking at the entire event sample \cite{Cheng:2008mg}. In the case of two visible particles, one can restrict the masses of all new particles to an allowed mass region \cite{Cheng:2007xv}.  As pointed out in \cite{Cheng:2007xv}, the correct mass values are typically near the boundary of the allowed region. Furthermore, if $N_{solns}(m_{LSP})$ is the number of events consistent with an LSP mass $m_{LSP}$, its true value can be  estimated by identifying where $dN_{solns}/dm_{LSP}$ changes suddenly.  
Another technique utilizes the $M_{T2}$ variable \cite{Lester:1999tx,Barr:2003rg}, whose definition requires a trial mass for the missing energy particle, $m_{LSP}$.  As noticed in \cite{Cho:2007qv,Gripaios:2007is, Barr:2007hy,Cho:2007dh}, for cascade decays where some of the intermediate state heavy  particles are offshell, there is a kink for ${\rm max} \left[M_{T2}(m_{LSP})\right]$ at the true mass of the LSP (the kink is also present, but less pronounced, when the intermediate states are onshell).  Recently, the connection between these two methods was discussed \cite{Cheng:2008hk}. It was also recently shown that an extension of the $M_{T2}$ method can be used to determine all particle masses in a given cascade decay \cite{Burns:2008va}.
  
We can now describe how these methods apply to our different FDM scenarios. In the case of SUSY with non-zero $LLE$ couplings, some SUSY cascade decays have already been studied.  The case of neutralino production followed by leptonic decays, {\it i.e.}, $\chi_1^0 \to l \bar{l}' \nu$, has been analyzed for the opposite sign, same flavor combination  when the neutralino decay is mediated by an onshell slepton \cite{Cheng:2007xv}.  While the signal reconstruction is potentially clean, the expected number of events is small unless $\chi_1^0$ is light and Higgsino-like. The possibility that the two leptons have different flavors complicates the event reconstruction but allows one to look for events with  $2\mu^{\pm} 2e^{\mp} + E_T^{\rm miss}$, which are very hard to obtain out of background processes.  At any rate, with enough events, the LSP mass can be determined by the kinematic method.

In the case where the decay occurs directly to a three-body final state ({\it i.e.}, the charged slepton mediating the decay is offshell), the kink method \cite{Cho:2007qv, Gripaios:2007is, Barr:2007hy,Cho:2007dh} can be used. In this case, the endpoint of $M_{T2}$ as a function of $m_{LSP}$ is predicted to be linear all the way to zero mass.  If the collider stable particle is instead massive, the linear behavior would change to the form $c_1 + \sqrt{c_2 + m_{LSP}^2}$ at the true value of the mass of the collider-stable state, giving a kink between the two functional forms.  Another useful topology with a more substantial cross section is squark production followed by cascade-decays into charged-leptons and neutrinos, including $\tilde{q}\to q \chi_1^0 \to q l \bar{l'} \nu$.  Such decays were explored in \cite{Cheng:2008hk}, where either kinematic \cite{Cheng:2008mg} or $M_{T2}$ techniques can be applied when the $\chi_1^0$ decay is respectively onshell or offshell.  

In the case of SUSY with non-zero $LQD$ couplings, leptons are no longer guaranteed to appear in the final state, so it is prudent to focus on hadronic modes.  For concreteness, we focus on the case where gluinos are pair-produced and each decays into two jets (one of them a $b$-jet) and a neutrino, Eq.~(\ref{gluino_LSP}).  The decay is mediated by an onshell or offshell sbottom.  This is similar to the case of the dilepton cascade mentioned above, except for the presence of combinatorial complications related to pairing the different jets.  The kink method can be applied to both cases \cite{Cho:2007qv,Gripaios:2007is, Barr:2007hy, Cho:2007dh}, where the kink is more prominent when the sbottom is offshell.

The accuracy of the different mass-reconstruction techniques is still an open question, so it is not yet clear to what level one can differentiate a massive invisible particle from a massless one.  We anticipate that the accuracy will depend on the decay mode and the masses of all parent particles. Naively, one would presume that a 50 GeV LSP will not be confused with a neutrino, while a 5 GeV LSP will appear, as far as the collider experiments are concerned, massless. 

A detailed discussion of this crucial issue -- how heavy must the LSP be so that we can tell it is not massless -- is beyond the scope of this paper. We would, however, like to discuss a couple of simple-minded estimates. We first use ``data'' from \cite{Cho:2008cu}, associated to a dilepton $t\bar{t}$ sample. In this case, the $M_{T2}$ kink method can be used to measure the neutrino mass.  From a $\chi^2$ fit to the kink, we obtain, at the 90\% confidence level, $m_{\nu}<17.5$~GeV. We also looked at the technique of subsystem $M_{T2}$, advertised in \cite{Burns:2008va}. Here one can completely reconstruct the masses of all ``unknown'' states (the top-quark, the $W$-boson and the neutrino) as a function of the endpoints of three different kinematic distributions involving the visible particles and the missing transverse energy \cite{Burns:2008va}. We estimate that if the $E_{ijk}$ variables ($i,j,k=0,1,2$, see \cite{Burns:2008va}) can be measured with an uncertainty of $\pm 1$~GeV (equivalent to a percent-level measurement), $m_{\nu}$ is constrained to be less than 28~GeV.  If we repeat the analysis for a $400$~GeV ``top'' that decays into a 200~GeV ``$W$-boson'', the uncertainty on the mass of the massless ``neutrino'' is, not surprisingly, larger:  $m_{\nu}<43$~GeV.  Techniques that are able to directly solve for the LSP mass \cite{Cheng:2008mg} potentially  have smaller uncertainties.
However, these naive estimates seem reasonable and suggest that the mass of a massless invisible particle can only be constrained to be less than $O(10)$~GeV.  

\section{Conclusions} \label{sec:conclusion}

By combining various experimental signals, the expectation is that we will soon have a concrete understanding of the nature of dark matter.  In the case of the WIMP scenario, DM will be produced at the LHC and will manifest itself in non-SM-like events with leptons and/or jets plus missing transverse energy, $E_T^{\rm miss}$.  Given, however, the enormous bias in favor of this possibility, it is prudent to consider explanations for $E_T^{\rm miss}$ events which do not involve the existence of new collider stable particles.  In this paper, we discussed an alternative scenario where the missing energy is solely due to the well-known standard model source of missing energy, neutrinos.     

We've focused on `fake dark matter' scenarios, where it is tricky to identify that neutrinos are the source of the the missing energy.  This required us to consider extensions of the standard model where neutrinos are produced in the cascade decays of new particles, while there are no or few new physics events with no missing energy. We also chose scenarios where one does not expect to observe displaced decay vertices.  Under these constraints, many models of fake dark matter exist; we focused on a simple leptoquark example and on supersymmetric models with trilinear RPV.  To our knowledge, the fact that several SUSY models with RPV can mimic dark matter signals at colliders was not appreciated in the literature.

We've presented an initial foray into collider analyses that could distinguish fake dark matter from collider-stable WIMPs.  Model-dependent tests exist.  In the examples presented, there are often particle tags for each cascade decay (leptons, bottom jets, $W$, $Z$ and Higgs bosons).  Observing events without some of these particle tags would disfavor different fake dark matter scenarios, while the presence of events consistent with specific tags would highlight scenarios about which one should worry.  The supersymmetric FDM models have phenomenology similar to low-scale mediated supersymmetry breaking models.  In some cases, FDM may be distinguished by its lack of displaced vertices and flavor violating particle tags.  In the $LLE$ model, for example, the particle tags can violate lepton flavor, which can be identified given a large enough event sample.  

As a model-independent test, we also considered the prospect of measuring the mass $m_{LSP}$ of the collider-stable particle through both the kinematic methods \cite{Kawagoe:2004rz,Cheng:2007xv,Cheng:2008mg} and the $M_{T2}$ kink method \cite{Cho:2007qv, Gripaios:2007is, Barr:2007hy,Cho:2007dh}.  The kinematic methods apply well to final states containing leptons, produced, say, in models with $LLE$ RPV, where kinematic fitting is not complicated by combinatorics.  On the other hand, the kink method is applicable to the all-hadronic decays associated to gluino pair-production in the $LQD$ RPV case.  We also presented rough estimates of the mass resolution of such methods, which was of the order of tens of  GeV.  Although naive, this resolution suggests that mass measurements are only capable of unambiguously distinguishing neutrinos from heavy weak-scale WIMPs.    

There are many possible directions for future research.  It should be possible to build models of fake dark matter starting from other theories containing a WIMP and violating the symmetry that prevents the WIMP decay.  In particular, scenarios with one or more universal extra dimensions and little higgs models with $T$-parity should have their own associated fake dark matter manifestations that may have other distinguishing features.  One can also relax (some of) our requirements for fake dark matter -- no new physics events without $E_T^{\rm miss}$ and no displaced decay vertices -- which could provide other interesting examples.  There are also many possibilities in confirming and extending the different collider analyses that were only briefly considered here.                  

In summary, the fake dark matter scenario is an example of the complexity behind interpreting future LHC results.  Despite many arguments for revolutionary discoveries like dark matter being tied to new physics events, it is important to consider alternative explanations.  If possible, LHC data should be analyzed without these priors in order to prevent the naive pursuit of red herrings from clouding a clear understanding of the physics surrounding the weak scale.

\section*{Acknowledgements} 
SC thanks David E. Kaplan, Markus Luty, and Jay Wacker for useful conversations, Neal Weiner for both useful conversations and endless encouragement, and Sara Mader for inspiration.  
AdG thanks the hospitality of the Physics Departments at Columbia University and NYU, where parts of this work were initiated, and of the Fermilab Theory Division, where this work was concluded.  SC  acknowledges the hospitality and support of the Kavli Institute for Theoretical Physics China, CAS, Beijing 100190, China, where some of this work was undertaken.  We are also grateful to Zhenyu Han and KC Kong for very useful conversations regarding mass measurements at colliders and comments on the manuscript.
The work of SC was supported by NSF CAREER grant PHY-0449818 and DOE grant \# DE-FG02-06ER41417. AdG is sponsored in part by the US Department of Energy Contract DE-FG02-91ER40684.

\bibliography{fakeDM}
\bibliographystyle{apsrev}

\end{document}